\documentstyle[12pt]{article}
\def\C{{\bf 126}}
\def\X{{\bf 10}}
\def\T#1{$\times 10^{#1}$}
\newcommand{\nummer}[1]{\baselineskip=14pt\hskip 12 cm #1 \par}
\newcommand{\datum}[1]{\hskip 12 cm #1}
\newcommand{\titel}[1]{\Large \vskip 3 true cm\begin{center}#1\end{center}
            \normalsize\vskip 1.0 true cm}
\newcommand{\autor}[1]{\normalsize\begin{center}#1\end{center}\vskip 0.5cm}
\newcommand{\adresse}[1]{\begin{center}#1\end{center}\vskip 2 true cm}
\renewcommand{\abstract}[1]{\hfil\parbox{15.4 true cm}
{\Large ABSTRACT: \normalsize #1} \hfil \vfil \Large
\normalsize\eject}
\begin{document}
\begin{titlepage}
\nummer{WU B 94-44}
\datum{Feb.  \quad 1995}

\titel{\bf Predicting the neutrino-spectrum in SUSY-SO(10)}
\autor{ YOAV ACHIMAN~\footnote{e-mail: achiman@wpts0.physik.uni-wuppertal.de}
\quad {\em and} \quad THORSTEN GREINER~\footnote{e-mail:
greiner@wpts0.physik.uni-wuppertal.de}}
\adresse{Department of Physics \\
         University of Wuppertal \\
         Gau\ss{}str. 20, D-42097 Wuppertal \\
         Germany}
\abstract{We present a systematic search for SUSY-SO(10) models which
predict the neutrino properties. The models are based on the five sets
of quark mass matrices, with texture zeros, discussed recently by Roberts,
Ramond and Ross. We found 8 such neutrino textures three of which
can solve the solar neutrino problem. The latter have tau-neutrino masses
of few eV i.e. relevant for cosmology and $\nu_{\tau} - \nu_{\mu}$ mixing
angles that can be observed by the CHORUS, NOMAD and P803 experiments.
}
\end{titlepage}
\textwidth=15.0 true cm
\textheight=22.0 true cm
\voffset -2.0cm
\hoffset -1.0cm
\parindent 0pt
\parskip 14pt
\hfuzz 1cm
\newcommand{\bye}{\end{document}}
\newcommand{\ns}{\vfill\eject}
\newcommand{\be}{\begin{equation}}
\newcommand{\ee}{\end{equation}}
\newcommand{\F}{Fritzsch\ }
\newcommand{\FT}{Fritzsch texture\ }
\newcommand{\gsim}{\lower.7ex\hbox{$\;\stackrel{\textstyle>}{\sim}\;$}}
\newcommand{\lsim}{\lower.7ex\hbox{$\;\stackrel{\textstyle<}{\sim}\;$}}
\def\NPB#1#2#3{Nucl. Phys. B {\bf#1} (19#2) #3}
\def\PLB#1#2#3{Phys. Lett. B {\bf#1} (19#2) #3}
\def\PLBold#1#2#3{Phys. Lett. {\bf#1B} (19#2) #3}
\def\PRD#1#2#3{Phys. Rev. D {\bf#1} (19#2) #3}
\def\PRL#1#2#3{Phys. Rev. Lett. {\bf#1} (19#2) #3}
\def\PRT#1#2#3{Phys. Rep. {\bf#1} C (19#2) #3}
\def\ARAA#1#2#3{Ann. Rev. Astron. Astrophys. {\bf#1} (19#2) #3}
\def\ARNP#1#2#3{Ann. Rev. Nucl. Part. Sci. {\bf#1} (19#2) #3}
\def\MODA#1#2#3{Mod. Phys. Lett. A {\bf#1} (19#2) #3}
\def\NC#1#2#3{Nuovo Cim. {#1} (19#2)#3}
\def\Platz{\rule[-6pt]{0pt}{24pt}}
\def\pa{\ $\phi_{\bf 10}$\ }
\def\pai{\ $\phi_{\bf 10}^{ij}$\ }
\def\pb{\ $\phi_{\bf \overline{126}}$ \ }
\def\pbi{\  $\phi_{\bf \overline{126}}^{ij}$ \ }
\def\mr{ \ $M_{\nu_R}$ \ }
\def\hu{ \  $H_u$ \ }
\def\hd{ \  $H_d$ \ }
\def\mnd{\ $M_{\nu_D}$ \ }
\def\mr{\ $M_{\nu_R}$ \ }
\section{Introduction}

The recent strong evidence for the top quark in CDF~\cite{cdf} completed the
information we have on the masses and mixing angles of the quarks.
It emphasizes, however, at the same time our ignorance of their origin.
As the fermionic masses are free
parameters in the standard model (SM), an embedding into a grand-
unified-theory~(GUT) can help. This is also
suggested by the unification of the gauge coupling constants of the
SM~\cite{sun}, at $10^{16}$ GeV, provided the spectrum is extended
into that of the minimal SUSY-SM (MSSM)~\cite{mssm}.\footnote{ Another
possibility is to
introduce an intermediate breaking scale at $\approx 10^{12}$GeV.~\cite{int} }
GUTs give relations between the
Yukawa coupling constants of different flavours, like the successful\quad
$Y_{\tau}(GUT)\simeq Y_b(GUT)$\quad
lepton-quark one. Yet, the complete understanding of the mass-mixing
pattern requires
relations between the families. This can come only from outside
the GUT, by using a family-symmetry (or superstrings?).
The only phenomenological
indication in this direction is that the mixing angles and masses
of the quarks are consistent with the appearance of texture zeros in the
Yukawa matrices~\cite{tex}.

A recent study by Roberts, Ramond and Ross~(RRR)~\cite{rrr} found five
different sets of symmetric quark mass matrices with texture zeros,
which account for the quark masses and mixing. Special examples,
like the Fritzsch~\cite{fr} texture, where known before. Also Dimopoulos,
Hall and Rabi~(DHR)~\cite{dhr} discussed in detail quark and charged leptons
mass matrices
suggested by Harvey, Ramond and Reiss~\cite{hrr}, in terms of a SUSY-
SO(10) broken directly into the MSSM.

All this is true for the quarks and charged leptons.
The neutrino-masses and mixing
are completely unknown. Except for possible experimental indications coming
from the solar-neutrino-puzzle~(SNP)~\cite{snp},
the depletion of the atmospheric $\nu_{\mu}$~\cite{atm} and some
cosmological dark matter arguments~\cite{cos}. All of which are consistent
with possible neutrino
masses in the range of $10^{-5}\mbox{ eV}  - 3\mbox{ eV}$ .\\
Such small neutrino masses are obtained in in L-R symmetric GUTs, like
SO(10), using the see-saw mechanism. This means that the SU(5) singlet
RH-neutrinos acquire large Majorana masses. The diagonalization of the
complete neutrino mass matrix leads then to three small eigenvalues.

In a previous paper~\cite{ag}, we were able to predict the neutrino
properties, by requiring that all matrices, including the
RH-neutrino Majorana mass one, have the same Fritzsch-texture.
The model was based on SUSY-SO(10) with the scale of the RH-neutrino
mass matrix taken at the unification energy -- as is natural in
SUSY theories. It gives
neutrino masses and mixing angles consistent with a possible
solution of the solar neutrino puzzle, without the need for a free parameter.
Unfortunately, if top was
observed at CDF, its mass is too high to be consistent with such a model.

In order to be able to predict the neutrino properties in terms of more
complicate textures, we must use stronger assumptions.
The RH-neutrino scale becomes then a free parameter and to solve  the SNP, it
must be lower then the GUT scale by several orders of magnitude.
Assuming, as in almost all recent fermionic mass models,
that the SUSY-GUT is broken directly into the MSSM, it is not
clear where this intermediate mass scale is coming from.

In this paper we use SUSY-SO(10) with
the three families of the quarks and the leptons,
including $\nu_R$, in the ${\bf 16}$ representation $\Psi_i$,
$i=1,2,3$. In the view of the content of,
$$
{\bf 16}\times{\bf 16}=({\bf 10}+{\bf 126})_{symmetric}+{\bf 120}_{antisymm.}
$$
only the \pa and \pb
Higgs representations can contribute to the symmetric Yukawa terms.\\
The most general Yukawa Lagrangian at the GUT scale is then:

\be
{\cal L}_Y = \sum  \overline{\Psi_i}^c \Psi_j (Y_{ij}^{\bf 10}
\phi_{\bf 10}^{ij}  +
Y^{\bf \overline{126}}_{ij}\phi_{\bf \overline{126}}^{ij}) \ .
\ee

Note, that one can absorb the difference between \quad $Y_{ij}^{\bf 10}$
\quad and \quad $Y^{\bf \overline{126}}_{ij}$ in the VEVs, and use only one
effective Yukawa matrix, if all the Higgs representations are different.
(This is was used in the previous paper \cite{ag} but is not true here,
as we shall see later).

Our aim is to predict the neutrino properties in terms of the
mass matrices of the quarks and the charged leptons. In order to do this
we shall use the
requirements suggested by DHR~\cite{dhr}\footnote{They used those requirements
for ``their'' texture, which is very probably excluded experimentally as it
requires $|V_{cb}|>0.5$.} and apply them to the five
texture sets of RRR.

The requirements combine actually {\em predictibility} and {\em minimality}
as follows:
\begin{enumerate}
\item{The textures of the mass matrices are dictated by discrete
symmetries and the directions of the VEVs in such a way that the
minimal number of higgs multiplets is used.}
\item {Each fermion mass  matrix element is generated by a VEV of only one
of the \pa or \pb multiplets.}
\item {All entries of the RH-neutrino Majorana mass matrix, \mr,
must be induced by one \pb
multiplet and in such a way the matrix is not singular.}
\end{enumerate}

The textures of RRR do not tell us if the non-vanishing entries are due to
the \pa or the \pb Higgs representation.
 More information about $M_d$, the down quarks mass matrix, can
be obtained using the ``connection'' between this matrix  and that of the
charged leptons, $M_\ell$.
In view of the fact that the \pb contributions come with a relative
Clebsch-Gordan coefficient of (-3), the fit of the $M_\ell$ elements to
the lepton masses can tell us where \pb contributes.
It was already pointed out by Georgi and Jarlskog~\cite{gj} that the Ansatz
$$
m_{\tau} = m_b \qquad m_{\mu} = 3m_s \qquad m_e = 1/3 m_d
$$
at the GUT scale, works very well. This Ansatz can be generated by a factor
(-3) in the \ $(M_\ell)_{22}$ \ matrix element~\cite{hrr}~\cite{dhr}. RRR
checked it  for their textures in terms of the SUSY-GUT broken directly
into the MSSM~\cite{rrr}.  This is obviously consistent with our
requirement (1). Note, also that the leading (3,3) elements are generated
in such a case
by \pa. Thus, they obey  \ $(M_d)_{33} = (M_{\ell})_{33}$, and one has
in addition, the successful approximate (Yukawa) \quad
$Y_b(GUT) \simeq Y_\tau(GUT)$ \quad unification.

The structure of the mass matrix of the ``up"-quarks, $M_u$, cannot be fixed
using similar arguments, as the
related neutrino Dirac mass matrix, $M_{\nu_D}$, is
phenomenologically unknown. However, we will
show that our minimality and predictibility requirements limit
considerably the possibilities.

We found using our method eight sets of symmetric textures which
predict the neutrino properties -- up to the overall mass scale of the RH-
neutrino masses. \\
We evolved then the Yukawa matrices, from the GUT scale to low energies, using
the renormalization group equations (RGEs). The resulting matrices are then
fitted
to the low-energy  experimental data, and this fixes the quark parameters at
the GUT scale.
Those parameters dictate the entries of the light-neutrino parameters, for each
one of the eight sets of textures. At the same time, we have also
predictions for certain quantities in the quark-sector which we can
used as a test. The light (see-saw) neutrino mass matrix is then evolved
to low energies.

The resulting neutrino properties are given in terms of their mass ratios
and mixing angles. The absolute neutrino masses can be obtained only
when the intermediate scale, relevant for the overall scale of the RH-
neutrino mass matrix, is given.

The mixing angles of all the eight texture sets are such that
\quad ${\sin^2}2\theta < 0.2 $,\quad
and hence, we cannot have vacuum oscillation as a solution to the solar
neutrino puzzle. Also, the possible depletion of the atmospheric
$\nu_{\mu}$
cannot be accounted for. The values of the $\nu_e-\nu_{\mu}$ mixing angle
(i.e. \quad ${\sin^2}2\theta_{e\mu}$ \quad)
are generally in
the range of the small angle (i.e adiabatic) MSW~\cite{msw} solution
to the SNP. Requiring that  \quad ${\Delta m}_{e\mu}^2$ \quad
has the right value for this solution, we obtain for the RH-neutrino scale:
$$
M_R \sim 10^{13} - 10^{14}  GeV.
$$
The corresponding masses of $\nu_\tau$ are of few eV  -- in
the range interesting for cosmology~\cite{cos}. At the same time the
values of \quad ${\sin^2}2\theta_{\mu\tau}$ \quad are such that $\nu_\tau$
oscillations
will be observed in experiments like CHORUS~\cite{chorus}, NOMAD~\cite
{nomad} and P803~\cite{p803}.

The plan of the paper is as follows. In sect.~2 the models (and their
discrete symmetries) will be discussed in detail.
Sect.~3 will explain the details of our numerical analysis.
In sect.~4 we will give and discuss the results in the neutrino sector.
Conclusions and remarks can be found in sect.~5.

\section{The models}

The general Yukawa Lagrangian is given in eq.(1). To have the
actual form of the mass matrices, one must give the Yukava coupling
constants $Y_{ij}^{\bf 10}$ and $Y_{ij}^{\bf \overline {126}}$ and the
VEVs of the Higgs representations \pai and \pbi. The discrete symmetries,
to be discussed later, will play here an important role. Those symmetries
fix the non-zero entries of the Yukawa matrices and ensure the stability
of our predictions.

Let us, however, discuss first the possible entries to the matrices on
a pure phenomenological level.
Both \pa and \pb can develop VEVs in the directions of the down and/or
the up quarks. However, only the \pb multiplets allow for a B--L
violating VEV, which generates the Majorana mass matrix of the RH-
neutrino. We assume, as usual, that below the GUT scale one has
effectively the MSSM with two doublets of Higgs \hu and \hd. These
are mixtures of the SM doublet components of all scalar SO(10)
representations,
also those needed for the local symmetry breaking. We can therefore
separate the Yukawa terms into five groups, even at the SO(10) GUT
scale as follows:

\renewcommand{\arraystretch}{1.5}
\be
\begin{array}{ccc}
{\cal L}_Y & = &\sum Y_{ij} \left \{ a^{ij} \left[ ( \overline{d_{Ri}} d_{Lj} +
\overline{\ell_{Ri}} \ell_{Lj}) H_{{\bf 10},d}^{ij}
  + (\overline{u_{Ri}}{u_{Lj}} + \overline{\nu_{Ri}}\nu_{Lj}) H_{{\bf
10},u}^{ij}
\right] \right .\\
  & + & b^{ij} \left [ (\overline{d_{Ri}} d_{Lj} - 3
\overline{\ell_{Ri}}{\ell_{Lj}})
   H_{{\bf 126},d}^{ij} + (\overline{u_{Ri}} u_{Lj} - 3
\overline{\nu_{Ri}}{\nu_{Lj}})
    H_{{\bf 126},u}^{ij} \right . \\
 & + & \left .\left .\overline{\nu_{Ri}} \nu_{Rj} (-3)
\phi_{{\bf {\overline 126},1}_{SU(5)}}^{ij} \right ]
 \right \}\ .
\end{array}
\ee

In view of the requirement (2) that only one of the \pai or \pbi can contribute
to the mass matrices, we have for the non-vanishing Yukawa matrix elements
only one of the two possibilities:
$$
(a^{ij} , b^{ij})= (1,0)  {\rm\quad or \quad}  (0,1).
$$
The quark mass matrices develop below the GUT scale, in terms of the MSSM,
the following contributions, which define the effective Yukawa matrices
used in the RGEs:

\be
(M_d)_{ij} = Y_{ij} (a_{ij}{\gamma_d}^{ij} + b_{ij}{\theta_d}^{ij})
\cos\beta\ \upsilon
\ee

\be
(M_u)_{ij} = Y_{ij} (a_{ij}{\gamma_u}^{ij} + b_{ij}{\theta_u}^{ij})
\sin\beta\ \upsilon
\ee

where, $\gamma^{ij}$ and  $\theta^{ij}$ account for the amount of
mixing of the VEVs of the MSSM doublets $<H_d>$ and $<H_u>$ . Also,
as usual in the MSSM:

$$
\tan \beta = \frac{<H_u>}{<H_d>} {\rm\qquad and\qquad}
 \upsilon = \sqrt{{<H_u>}^2 + {<H_d>}^2} = 174\mbox{ GeV}\ .
 $$

Now, for the phenomenological ``good'' textures, there are additional
restrictions:\\
The Yukawa couplings \quad  $Y_{ij}$ \quad vanish when the corresponding
texture
zeros are common to both $M_d$ and $M_u$.
E.g\quad $Y_{ij}=0$\quad in all texture sets.\\
For zero entries in only one matrix we have:
$$
a_{ij}{\gamma}_u^{ij} + b_{ij}{\theta}_u^{ij} =0  \qquad or \qquad
a_{ij}{\gamma}_d^{ij} + b_{ij}{\theta}_d^{ij} =0 .
$$
As for the non-vanishing $(i,j)$ matrix elements -- it is impossible
to say which Higgs representation \pai or \pbi contributes, as long as only
the quark masses and mixing angels are used.

Our phenomenological discussion is based on the five sets of texture zeros
for the quarks, given in table \ref{SQM}.

Now, to predict the neutrino matrices we must know which of the Higgs
representations,  \pa or \pb, contributes to the different matrix elements.
As was already discussed in the introduction, the structure of $M_d$
and $M_{\ell}$ is fixed by the need to have the approximate Yukawa
unification and the Georgi-Jarlskog mass relations. The result is that
all the non-vanishing matrix elements will be generated by \pa~,\linebreak[3]
except for the (2,2) one which is due to \pb . (I.e. it obtains  a relative
factor (-3) in \  $(M_{\ell})_{22}$ \ relative to  \ $(M_d)_{22}$ ).\\
The explicit structure of those matrices , for the different textures,
can be found in table \ref{STRUCT}.

In order to fix the structure of $M_u$ we must go in a different direction,
as $M_{\nu_D}$ is unknown phenomenology. We shall use our
predictability and minimality requirements to restrict considerably
the number of possibilities. The resulting \  $M_u$ \ textures  will dictate
the neutrino matrices.\\
The arguments go as follows:\\
All non-vanishing entries to \quad $M_{\nu_R}$ \quad must be generated by
one \pb and
this should be induced via our discrete symmetries. Those, however,
allow for one Higgs multiplet to couple to at most two $(i,j)$ entries.
Hence, only the following non-singular possibilities are open:\\
\renewcommand\arraystretch{1.0}
$$
 M_{\nu_R}^I =  \left (\begin{array}{ccc}
0 & y & 0 \\
y & 0 & 0 \\
0 & 0 & x
\end{array} \right )\quad , \quad
 M_{\nu_R}^{II} =    \left (\begin{array}{ccc}
0 & 0 & y \\
0 & x & 0 \\
y & 0 & 0
\end{array} \right )
$$
and
$$
 M_{\nu_R}^{III} =   \left (\begin{array}{ccc}
x & 0 & 0 \\
0 & 0 & y \\
0 & y & 0
\end{array} \right ) .
$$
The two last possibilities, however, cannot be realized in our textures.
For $M_{\nu_R}^{III}$ it is clear because in all the five quark texture \quad
$Y_{11} = 0$. For $M_{\nu_R}^{II}$ it is more complicated. If one Higgs
representation induces contributions to two entries (i,j) and (k,l) in
one of the matrices, it means that $\overline {\Psi_i^c} \Psi_j$ and
$\overline {\Psi_k^c}\Psi_l$ have the same quantum numbers. Thus, {\em all}
mass matrices acquire a contribution in {\em both} entries or {\em no one at
all}.
(In the last case, all Higgs representations with the above quantum namber
do not develop a VEV in the relevant direction).
In our case \ $(M_d)_{22} \not=0$  \  while  \ $(M_d)_{13} =
(M_d)_{31} = 0 $ \ in all textures and  \  $M_{\nu_R}^{II}$ \
is inconsistent with the above requirement.

Now, \pb which generates \  $ M_{\nu_R}^I$ \  can contribute to \ $M_u$ \
and
 \ $M_{\nu_R}$ \ only,  as \quad $(M_d)_{33}$ \quad must come from a \pa .
Hence, the \pb representation which induces the Majorana $M_{\nu_R}$
can contribute to $M_u$ and $M_{\nu_D}$ only. Minimality then requires
that this must be the case.

One finds, by explicit observation, that \  $M_{\nu_R}^I$ \ is relevant
for the texture sets 1,2 and 4.
This fixes three matrix elements in \  $M_u$.
The other entries required by the five quark textures can get
contributions from both \pa or \pb.\  $M_{\nu_D}$ \ is then obtained from
$M_u$ using suitable Clebsch-Gordan factors. Note, that the magnitude
of those contributions is given by the phenomenology i.e by fitting
the evolved $M_u$ matrix to the observed masses and mixing angles. Only the
factors accompanying these contributions in \  $M_{\nu_D}$ \  are
dictated by
the choice of \pa or \pb.

We have, therefore, several possible combinations for each texture set and
in total eight  different ones. Those are presented in table \ref{STRUCT}.

Once the neutrino matrices \mnd and \mr are known, we can construct
the see-saw matrix for each model, in the form:

\be
M_{\nu}^{light} \simeq - M_{\nu_D} M_{\nu_R}^{-1} M_{\nu_D} .
\ee

After this matrix is evolved to low energies (see next section) it gives us
the neutrino masses and mixing angles. To calculate the mixing angles one must
obviously consider the charged lepton mass matrix as well. The angles
are, however, independent on the overall mass scale of the RH-neutrinos.
The latter  is a free parameter in our models and hence, we
predict the neutrino mass ratios only. The RH-neutrino scale will be fixed
latter for models with mixing angles which allow for a solution to the
SNP, such that \  $\Delta m_{e\mu}^2$ \ have the right value.

We know now phenomenologically what the different textures are and it remains
only to show how those textures can be induced using discrete symmetries.
This is actually strait forward and very similar in the
different models.

Let the fermions and Higgs representation have the following
transformation properties under our symmetry:
$$
\begin{array}{ccc}
\psi_j \rightarrow e^{i\alpha_j}\psi_j & \hbox{ and } &
\phi_j^{\bf 10} \rightarrow  e^{i\beta_j} \phi_j^{\bf 10} \\[10pt]

 &  & \phi_j^{\bf\overline{126}} \rightarrow e^{i\gamma_j}
\phi_j^{\bf\overline{126}} .
\end{array}
$$
We must require that \ $(M_u)_{12}$ \ and \ $(M_d)_{33}$ \ are generated by
one \pb. Hence,\\
$$
{\alpha_1} + {\alpha_2} = 2{\alpha_3} = -{\gamma_1} .
$$
However, \ $M_d$ \ gets also contributions at the (1,2) and (3,3) entries,
via \pa. As our symmetry is on the SO(10) level, those
matrix elements also, must be due to the same Higgs representation i.e  \pa
in this case and \\
$$
\beta_1 = \gamma_1  .
$$

This means that \ $\phi_{\bf \overline {126}}^1$ \ generates a
light VEV in the u-direction while \ $\phi_{\bf 10}^1$ \ generates
one in the d-direction.
The other entries acquire contributions according to the corresponding
quantum numbers.\\

As an explicit realization we can take:\\
$$
\alpha_1 = 1 , \qquad \alpha_2 = 3  \qquad \hbox{ and } \qquad \alpha_3 = 2 .
$$
in this case:\\
$$
\beta_1 = \gamma_1 = -4 .
$$
E.g for the texture $1_I$ we have , in addition:\\
$$
\gamma_2(\phi_{\bf 10}^{22}) = \beta_2(\phi_{\bf \overline{126}}^{22}) =
 -2\alpha_2 = -6
$$
and
$$
\beta_3(\phi_{\bf 10}^{23}) =  -(\alpha_2 + \alpha_3) = -5  .
$$

So finally for this texture we need:\\
$$
3 \ \times \ \phi_{\bf 10}  \qquad and \qquad 2  \ \times  \
\phi_{\bf \overline{126}} \quad ,
$$

of which only \ $\phi_{\bf \overline{126}}^ 1$ \ generates a heavy VEV.

In all other models very similar discrete symmetries are needed.

\section{Renormalization Group Equations and Fits}

All the matrix elements of our matrices are in principle complex numbers.
One can, however, use the freedom to redefine the nine phases of the three
LH-doublets and six RH-singlets of the SM, to reduce considerably the
number of the ``physical" phases. In any case, symmetric quark matrices can be
always transformed into hermitian ones in this way~\cite{rrr}.
As we are interested only in the neutrino
sector and the leptonic phases cannot be observed - we use for simplicity
only one physical phase. Let us put it at the (1,2) matrix
element and in an hermitian way - as DHR~\cite{dhr} do.

As an example, we give in the following the explicit matrix elements
of the model
discussed in the previous section.

Model  $1_I$ :
\begin{equation}
{\bf Y}_U = \left(\begin{array}{ccc} 0 & C_u & 0 \\ C_u & B_u & 0\\ 0 & 0 & A_u
	          \end{array} \right)
\quad
{\bf Y}_D = \left(\begin{array}{ccc} 0 & D_d e^{i\phi} & 0 \\
				D_d e^{-i\phi} & C_d & B_d\\ 0 & B_d & A_d
	          \end{array} \right)
\end{equation}
\begin{equation}
{\bf Y}_{\nu_D} = \left(\begin{array}{ccc} 0 & -3 C_u & 0 \\ -3 C_u & B_u & 0\\
					0 & 0 & -3 A_u
	          \end{array} \right)
\quad
{\bf Y}_L = \left(\begin{array}{ccc} 0 & D_d e^{i\phi} & 0 \\
				D_d e^{-i\phi} & -3 C_d & B_d\\ 0 & B_d & A_d
	          \end{array} \right)
\end{equation}
\begin{equation}
{\bf Y}_{\nu_M} = \left(\begin{array}{ccc} 0 & C_u & 0 \\ C_u & 0 & 0\\
					0 & 0 & A_u
\end{array} \right)
\end{equation}

In order to extract the neutrino properties from these matrices, the matrix
elements $A_u$, $B_u$, \ldots, $D_d$ and $\phi$ have to be determined. We do
this, as usual, by fitting masses and mixing angles as predicted by the
textures to their experimental values. Since these specific textures for the
Yukawa matrices are given at the unification scale $M_X$, we must evolve the
matrices from the GUT scale, $M_X$ to low energies using the renormalization
group equations (RGEs) (see Appendix \ref{App1}).

In our model, the SUSY-SO(10) is broken at $M_X$ directly into the MSSM.
The MSSM is broken at the effective \ $M_{SUSY} \approx 100 GeV$ into
the SM which in broken in its turn effectively at $M_Z$ into \ $SU_C(3) \times
U_{EM}(1)$ .\
We take, as it is done in many papers, $M_{SUSY} = M_Z$. A different choice
will have only a minor effect on the neutrino properties.

The renormalization group equations for the gauge and Yukawa coupling
constants~\cite{rge} are coupled, non-linear first order differential
equations, which do not have a complete analytical solution.
We therefore use a numerical procedure both to solve the RGEs and to fit the
Yukawa matrix parameters.

For the Yukawa RGEs we use the effective Yukawa matrices $\Lambda^{u,d...}$
defined by
\be
M_u = \Lambda^u\ \upsilon \sin\beta \quad , \quad  M_d = \Lambda^d\ \upsilon
\cos\beta \quad  e.t.c.
\ee
using equations (3) and (4). Thus, for a given value of $\tan\beta$ ,
$\Lambda^i$ are obtained in terms of the mass matrices.
The explicit calculations
were done using the semi-analytic form due to Barger, Berger and
Ohmann~\cite{bbo} (see Appendix \ref{App1}). This form reduces the number of
variables significantly.
To fix the parameters of a given texture, we obtained  first the masses
and mixing angles as functions of the GUT scale parameters and then evolved
them to low energies. Those are then fitted to the experimental values using
the 'shooting' method~\cite{gauge1} (see Appendix \ref{App2}).

The run of the Yukawa and gauge coupling constants from $M_X$
down to $M_Z$ is done in terms of the
two loop RGEs of the MSSM~\cite{rge}.
The appropriate boundary conditions for the Yukawa and gauge couplings are
applied at $M_Z$. Below $M_Z$,
three loop QCD and one loop QED renormalization group equations, are used.
We compare then our parameters with the standard masses of the light quarks
and charged leptons given at $\mu = 1 GeV$, and the heavy quarks at
their physical masses (see Table \ref{DATA}).

Since all of our textures have eight parameters, we have to fit to eight
experimental quantities. We use, out of the  experimental data displayed
in Table \ref{DATA}
$m_b$, $m_c$, $m_u$, $m_e$, $m_\mu$, $m_\tau$,
$|V_{us}|$ and $|V_{cb}|$ to fix the texture parameters. In addition, we take
$\alpha_1(M_Z)$ and $\alpha_2(M_Z)$ to fix the GUT scale $M_X$ and the gauge
coupling $\alpha(M_X)$ at the GUT scale.

Now, the texture parameters found with the shooting procedure define the
see-saw matrix (5). We evolve this matrix from $M_X$ to $M_Z$ using
one-loop renormalization group equations \cite{SSRGE} (see Appendix
\ref{App3}).

As a result we obtain, for each texture, a set of solutions which give a
good fit to the data, i.e $\chi^2 < 1$ .
Those solutions are parametrized according to the value of $\tan\beta$.
One of the predicted parameters is $m_t$. The dependence of $m_t$ on
$\tan\beta$ is given in Fig.~1, for the texture $1_I$. For all other textures
it is practically the same, as we have in all our models the approximate
$\tau - b$ Yukawa unification. One sees , as it is already well known in this
case, that small and large values of $\tan\beta$ are preferred. As those
correspond to two different physical situations~\cite{tan}, we will
present our results for $\tan\beta = 1.5$ and $\tan\beta = 55$.

\section{Discussion of the results. }

Our results for the different textures are displayed in table \ref{RESULTS}.\\
Looking at this table one sees clearly that we do not have large mixing
angles. Practically speaking, all our solutions obey
$$
\sin^2 2\theta < 0.2 \ .
$$
This means that our models cannot allow for the
depletion of the atmospheric muon neutrinos~\cite{atm}. Also, vacuum
oscillations will not be able to serve as a solution to the SNP~\cite{snp}
and only the small angle (i.e adiabatic) MSW mechanism can work.

Using the recent estimate for the small angle MSW region~\cite{ks}:
$$
\sin^2 2\theta_{e\mu} =  6 \times 10^{-4} - 2 \times 10^{-2} ,
$$
one sees that the models
$ 1_{II}, \   4_{III}, \ 4_{IV}, \ $
can explain the SNP.
This is obviously provided~\footnote{Note, that this the mass difference
relevant for our mixing angles in the range $\sin^2 2\theta_{e\mu} \simeq
(1-2) \times 10^{-2}$ .}:
$$
\Delta m_{e\mu}^2  \simeq  4 \times 10^{-5} .
$$
This requirement fixes the RH-neutrino scale to be:
$$
M_R = 10^{13} - 10^{14} GeV .
$$
Knowing the neutrino mass ratios we can compute the corresponding masses of
the $\tau$-neutrinos. Those are found to be all in the few eV region i.e.
interesting for cosmology~\cite{cos}. Also, the corresponding $\nu_{\tau} -
\nu_{\mu}$ mixing angles are large enough for $1_{II}$ and $4_{IV}$
to be observed in the already runing
CERN CHORUS~\cite{chorus} and NOMAD~\cite{nomad} experiments, as well as in
the approved FERMILAB P803~\cite{p803} one.
Also, some of the models
which cannot solve the SNP have relatively
large $\sin^2 2\theta_{\mu\tau}$
which can be observed in the above experiments.

\section{Conclusions and remarks}

We looked in this paper for SUSY-SO(10) models which can predict the
neutrino-spectrum in terms of the ``known" parameters of the charged
fermions.
The main idea is to dictate the mass matrix of the RH-neutrinos rather than
simply  conjecture its form, as it is done in many models.
To do this we used the requirements of DHR and suitable discrete symmetries.
Starting then from the general classification of ``good"" symmetric
textures for the quark mass matrices by RRR, we predicted correspondingly
eight neutrino mass matrices at the GUT scale.
In evolving these mass matrices to low energies we made some approximations:
a)we neglected threshold effects at the GUT scale  as well as at
$M_{SUSY}$ which we took to be $M_Z$.
b)we started the renormalization of the see-saw matrix also from the
GUT scale and not from $M_R$.
c)we made a simplifying conjecture for the unobservable leptonic phases.
Those approximations, however,  cannot change the qualitative predictions
of our models. They can at most change somewhat the neutrino mixing angles.
Practically speaking, one must allow for up to 10\% deviations from our
predictions of the neutrino properties.

We also required that our SUSY-SO(10) is ``the whole story". I.e. we did not
use possible non-renormalizable effective contributions due to physics
at the Planck scale (like gravity or superstrings). Such contributions
are frequently used in recent papers. There are very many possible
contributions of which one picks up those suitable for his arguments and
neglects arbitrarily all others. Such a procedure destroys the predictibility
which is the main ingredient of our models. Also, one can imagine scenarios
where the non-renormalizable effects are negligible and that we actually
assume.

There is, however, one indirect evidence that physics at the Planck-mass
may be relevant to our models. This is related to the RH-neutrino mass-scale
which is a free parameter . Yet, to explain
the solar neutrino puzzle and get $\tau$-neutrino masses relevant for
cosmology, we need $M_R = 10^{13} - 10^{14} GeV$ which is equal to
$\frac{M_{GUT}^2}{M_{Planck}}$~. It is also intersting to
embed  such an intermediate scale into the local symmetry
breaking of SUSY-SO(10), in order to make it natural\cite{ag2}.

\begin{appendix}
\section{Appendix}
\subsection{Semi-analytic approach\label{App1}}
In the renormalization group equations for the Yukawa couplings the largest
Yukawa contributions come from the Yukawa couplings of the third generation
$y_t$, $y_b$ and $y_\tau$. In view of this fact, Barger et.~al.\ \cite{bbo}
find the following RGEs for the Yukawa couplings:
\begin{eqnarray}
{{d\lambda _i}\over {dt}}&=&{{\lambda _i}\over {16\pi ^2}}
\Bigg [x_1+x_2\lambda _i^2+
a_u\sum_ {\alpha}\lambda _{\alpha}^2|V_{i\alpha }|^2\nonumber \\
&&+{1\over {16\pi ^2}}\Bigg (x_3+x_4\lambda _i^2+x_5\lambda _i^4
+\sum _{\alpha}\Big (b_u\lambda _{\alpha}^2+c_u\lambda _{\alpha}^4
+(d_u+e_u)\lambda _i^2\lambda _{\alpha}^2\Big )
|V_{i\alpha }|^2\Bigg )\Bigg ]\;, \nonumber\\
& &\\
{{d\lambda _{\alpha}}\over {dt}}&=&{{\lambda _{\alpha}}\over {16\pi ^2}}
\Bigg [x_6
+x_7\lambda _{\alpha }^2+
a_d\sum_ i\lambda _i^2|V_{i\alpha }|^2\nonumber \\
&&+{1\over {16\pi ^2}}\Bigg (x_8+x_9\lambda _{\alpha }^2+x_{10}
\lambda _{\alpha }^4
+\sum _i\Big (b_d\lambda _i^2+c_d\lambda _i^4
+(d_d+e_d)\lambda _{\alpha }^2\lambda _i^2\Big )
|V_{i\alpha }|^2\Bigg )\Bigg ]\;, \nonumber\\
& &\\
{{d\lambda _a}\over {dt}}&=&{{\lambda _a}\over {16\pi ^2}}
\Bigg [x_{11}+x_{12}\lambda _a^2
+{1\over {16\pi ^2}}\Bigg (x_{13}+x_{14}\lambda _a^2+x_{15}\lambda _a^4
\Bigg )\Bigg ]
\;,
\end{eqnarray}
where $i=u,c,t$, $\alpha=d,s,b$ and $a=e,\mu,\tau$. The CKM matrix elements
$W_1=|V_{cb}|^2$, $|V_{ub}|^2$, $|V_{ts}|^2$, $|V_{td}|^2$, $J$ evolve
according to
\begin{equation}
{{dW_1}\over {dt}}=-{{W_1}\over {8\pi ^2}}
\Bigg [\left (a_d\hat{\lambda} _t^2
+a_u\hat{\lambda} _b^2\right )+{{1}\over {(16\pi ^2)}}(e_d+e_u)
\lambda _t^2\lambda _b^2\Bigg ] \;, \label{dW1dt}
\end{equation}
with
\begin{eqnarray}
\hat{\lambda}_b^2&=&\lambda _b^2\left (
1+{{b_u+c_u\lambda _b^2}
\over {16\pi ^2a_u}}\right )\;, \\
\hat{\lambda} _t^2&=&\lambda _t^2\left (
1+{{b_d+c_d\lambda _t^2}
\over {16\pi ^2a_d}}\right )\;.
\end{eqnarray}
For $W_2=|V_{us}|^2, |V_{cd}|^2, |V_{tb}|^2, |V_{cs}|^2, |V_{ud}|^2$ we have
\begin{equation}
{{dW_2}\over {dt}}=0\;. \label{dW2dt}
\end{equation}
The various coefficients for the RGEs in the MSSM are given in table
\ref{MSSMCoeff}.
\subsection{Determining the texture parameters\label{App2}}
To determine the texture parameters, we fit the low energy predictions for
masses and mixing angles to the experimental values. In our case, all textures
have eight parameters, so we fit to eight experimentally known quantities,
namely $m_b$, $m_c$, $m_u$, $m_e$, $m_\mu$, $m_\tau$, $|V_{us}|$ and
$|V_{cb}|$. The procedure employed here is called {\em `shooting'}.

Let $p_j$ be the texture parameters and $R_i(p_j)$ the low energy predictions
obtained by the above mentioned running procedure. Further, let $r_i$ denote
their experimental values. We then have to solve the system of equations
\begin{equation}
	R_i(p_j) - r_i = 0.
\end{equation}
This is done by an iterative numerical procedure.
\subsection{Renormalization of the see-saw matrix\label{App3}}
Following the authors of \cite{SSRGE}, we define the neutrino see-saw mass
coefficient
\begin{equation}
{1 \over 2} c_1^{ab} =
  c_{21}^{ab} =
  c_{22}^{ab} =
{1 \over 2} c_3^{ab} =
Y_l^{ca} (M_R^{-1})^{cd} Y_l^{db}.
\end{equation}

The renormalization group equations for the see-saw matrix in the strict SUSY
limit are:
\begin{eqnarray}
  {d \over dt} c^{ab}_1 &=& \frac{1}{16\pi^2}\Big(
  \left[{1\over2} 2g_1^2 + 2g_2^2
  + 6~tr\left(Y_uY^{\dagger}_u\right)\right] c^{ab}_1
  + \left(Y_l Y_l^\dagger\right)^{bc} c^{ca}_1
  + \left(Y_l Y_l^\dagger\right)^{ac} c^{cb}_1\nonumber \\
  & &-\left(2g_1^2 + 6g_2^2\right)
   \left(c^{ab}_{21} + c^{ba}_{21}\right)
  - \left(2g_1^2 + 6g_2^2\right)
  \left(c^{ab}_{22} + c^{ba}_{22}\right)\Big)
\end{eqnarray}
\begin{eqnarray}
  {d \over dt} c^{ab}_3 &=& \frac{1}{16\pi^2}\Big(\left[2 g_1^2+2 g^2_2
  +6~tr\left(Y_u Y^{\dagger}_u\right)\right]c^{ab}_3
  +\left(Y_lY^{\dagger}_l\right)^{bc} c^{ca}_3
  +\left(Y_l Y^{\dagger}_l\right)^{ac} c^{cb}_3\nonumber\\
  & & -\left(2~g_1^2 +6~g_2^2 \right)
  \left(c^{ab}_{21}+c^{ba}_{21}\right)
  -\left(2~g_1^2 +2~g_2^2 \right)
  \left(c^{ab}_{22}+c^{ba}_{22}\right)\Big)
\end{eqnarray}

\begin{eqnarray}
  {d \over dt} c^{ab}_{21} &=&\frac{1}{16\pi^2}\Big(\left[4~g^2_2 - 2~g_1^2
  +6~tr\left(Y_u Y^{\dagger}_u\right)\right]c^{ab}_{21}
  + 2~g^2_2 ~c^{ba}_{21}
  +\left(Y_l Y^{\dagger}_l\right)^{bc} c^{ca}_{21}
  +\left(Y_l Y^{\dagger}_l\right)^{ac} c^{cb}_{21}\nonumber \\
  & &+\left(g_1^2-g_2^2\right)\left(c_{22}^{ab}+c_{22}^{ba}\right)
  -{1\over 2}\left(g_1^2+5~g_2^2 \right)
  \left(c_1^{ab}+ c_3^{ab}\right)\Big)
\end{eqnarray}

\begin{eqnarray}
  {d \over dt} c^{ab}_{22}&=&\frac{1}{16\pi^2}\Big(\left[-4~g_1^2-2~g^2_1
  +6~tr\left(Y_u Y^{\dagger}_u\right)\right]c^{ab}_{22}
  +2~g^2_2~c^{ba}_{22}
  +\left(Y_l Y^{\dagger}_l\right)^{bc} c^{ca}_{22}
  +\left(Y_l Y^{\dagger}_l\right)^{ac}c^{cb}_{22}\nonumber \\
  & &+\left(g_1^2-g_2^2\right)\left(c_{21}^{ab}+c_{21}^{ba}\right)
  -4~g_2^2~c^{ab}_{21}
  -{1\over2}\left(g_1^2-g_2^2 \right)
  \left(c_1^{ab}+ c_3^{ab}\right)\Big)
\end{eqnarray}
\end{appendix}

\newpage
\section*{Tables}

\begin{table}[h]
\begin{center}
\begin{tabular}{|c|c|c|}
\hline
Texture & $Y_u$ & $Y_d$ \\
\hline
1 & $\left ( \begin{array}{ccc}
             0 & C & 0 \\
             C & B & 0 \\
             0  & 0  & A
             \end{array} \right )$ &
           $ \left ( \begin{array}{ccc}
             0 & F & 0 \\
             F^* & E & E' \\
             0  & E'  & D
             \end{array} \right )$ \\ [20pt]
2 & $\left ( \begin{array}{ccc}
             0 & C & 0 \\
             C & 0 & B \\
             0  & B  & A
             \end{array} \right )$ &
           $ \left ( \begin{array}{ccc}
             0 & F & 0 \\
             F^* & E & E' \\
             0  & E^{'*}  & D
             \end{array} \right )$ \\[20pt]
3 & $\left ( \begin{array}{ccc}
             0 & 0 & C \\
             0 & B & 0 \\
             C & 0 & A
             \end{array} \right )$ &
           $ \left ( \begin{array}{ccc}
             0 & F & 0 \\
             F^* & E & E' \\
             0  & E' & D
             \end{array} \right )$ \\[20pt]
4 & $\left ( \begin{array}{ccc}
             0 & C & 0 \\
             C & B & B' \\
             0 & B' & A
             \end{array} \right )$ &
           $ \left ( \begin{array}{ccc}
             0 & F & 0 \\
             F^* & E & 0 \\
             0  & 0  & D
             \end{array} \right )$ \\[20pt]
5 & $\left ( \begin{array}{ccc}
             0 & 0 & C \\
             0 & B & B' \\
             C & B' & A
             \end{array} \right )$ &
           $ \left ( \begin{array}{ccc}
             0 & F & 0 \\
             F^* & E & 0 \\
             0  & 0 & D
             \end{array} \right )$ \\
\hline
\end{tabular}
\end{center}
\caption{The five sets of symmetric quark mass matrices with texture zeros
found by RRR [6]}
\label{SQM}
\end{table}

\begin{table}[h]
\begin{center}
\begin{tabular}{|l|l|}
\hline
Gauge couplings \cite{gauge1,ckm} &
Quark masses    \cite{gl,ckm}\\
\hline
$\alpha_1(M_Z) = 0.01698\pm0.00009$ &
$m_u(\mbox{1 GeV}) = 5.1\pm 1.5 \mbox{ MeV}$\\
$\alpha_2(M_Z) = 0.03364\pm 0.0002$&
$m_d(\mbox{1 GeV}) = 8.9\pm 2.6 \mbox{ MeV}$\\
$\alpha_3(M_Z) = 0.120\pm 0.007 \pm 0.002$ &
$m_s(\mbox{1 GeV}) = 175 \pm 55 \mbox{ MeV}$\\
&
$m_c(m_c) = 1.27 \pm 0.05 \mbox{ GeV}$\\
&
$m_b(m_b) = 4.4 \pm 0.10 \mbox{ GeV}$\\
\hline\hline
Lepton masses   \cite{gauge1} &
CKM matrix entries \cite{ckm}\\
\hline
$m_e(\mbox{1 GeV}) = 0.496 \mbox{ MeV}$ &
$|V_{us}| = 0.218-0.224$\\
$m_\mu(\mbox{1 GeV}) = 104.57 \mbox{ MeV}$ &
$|V_{ub}| = 0.002-0.005$\\
$m_\tau(\mbox{1 GeV}) = 1.7835 \mbox{ GeV}$ &
$|V_{cb}| = 0.032-0.048$\\
\hline
\end{tabular}
\end{center}
\caption{Experimental data used to fix the GUT scale parameters}
\label{DATA}
\end{table}

\begin{table}[h]
\begin{center}
\begin{tabular}{|c|cc|}
\hline
Texture 1 & $\bf U$ & $\bf D$\\
\hline
I &
${\bf U} = \left(\begin{array}{ccc} 0 & \C_1 & 0\\ \C_1 & \X_2 & 0\\ 0 & 0 &
\C_1 \end{array}\right)$ &
${\bf D} = \left(\begin{array}{ccc} 0 & \X_1 & 0\\ \X_1 & \C_2 & \X_3\\ 0 &
\X_3 &
\X_1 \end{array}\right)$ \\
II &
${\bf U} = \left(\begin{array}{ccc} 0 & \C_1 & 0\\ \C_1 & \C_2 & 0\\ 0 & 0 &
\C_1 \end{array}\right)$ &
${\bf D} = \left(\begin{array}{ccc} 0 & \X_1 & 0\\ \X_1 & \C_2 & \X_2\\ 0 &
\X_2 &
\X_1 \end{array}\right)$\\
\hline
Texture 2 & & \\
\hline
I & ${\bf U} = \left(\begin{array}{ccc} 0 & \C_1 & 0\\ \C_1 & 0 & \X_2\\ 0 &
\X_2 &
\C_1 \end{array}\right)$ &
${\bf D} = \left(\begin{array}{ccc} 0 & \X_1 & 0\\ \X_1 & \C_2 & \X_2\\ 0 &
\X_2 &
\X_1 \end{array}\right)$\\
II & $ {\bf U} = \left(\begin{array}{ccc} 0 & \C_1 & 0\\ \C_1 & 0 & \C_3\\ 0 &
\C_3 &
\C_1 \end{array}\right)$ &
$ {\bf D} = \left(\begin{array}{ccc} 0 & \X_1 & 0\\ \X_1 & \C_2 & \X_2\\ 0 &
\X_2 &
\X_1 \end{array}\right)$\\
\hline
Texture 4 & $\bf U$ & $\bf D$\\
\hline
I & ${\bf U} = \left(\begin{array}{ccc} 0 & \C_1 & 0\\ \C_1 & \X_2 & \X_3 \\ 0
& \X_3 &
\C_1 \end{array}\right)$ &
$ {\bf D} = \left(\begin{array}{ccc} 0 & \X_1 & 0\\ \X_1 & \C_2 & 0 \\ 0 & 0 &
\X_1 \end{array}\right)$\\
II & ${\bf U} = \left(\begin{array}{ccc} 0 & \C_1 & 0\\ \C_1 & \C_2 & \X_2 \\ 0
& \X_2 &
\C_1 \end{array}\right)$ &
${\bf D} = \left(\begin{array}{ccc} 0 & \X_1 & 0\\ \X_1 & \C_2 & 0 \\ 0 & 0 &
\X_1 \end{array}\right)$\\
III & ${\bf U} = \left(\begin{array}{ccc} 0 & \C_1 & 0\\ \C_1 & \X_2 & \C_3 \\
0 & \C_3 &
\C_1 \end{array}\right)$ &
${\bf D} = \left(\begin{array}{ccc} 0 & \X_1 & 0\\ \X_1 & \C_2 & 0 \\ 0 & 0 &
\X_1 \end{array}\right)$\\
IV & ${\bf U} = \left(\begin{array}{ccc} 0 & \C_1 & 0\\ \C_1 & \C_2 & \C_3 \\ 0
& \C_3 &
\C_1 \end{array}\right)$ &
${\bf D} = \left(\begin{array}{ccc} 0 & \X_1 & 0\\ \X_1 & \C_2 & 0 \\ 0 & 0 &
\X_1 \end{array}\right)$\\
\hline
\end{tabular}
\end{center}
\caption{Explicit structure of the eight ``good'' textures}
\label{STRUCT}
\end{table}

\begin{table}[h]
\begin{center}
\begin{tabular}{|c|cc|cc|c|}
\hline
$\tan\beta$ & $\sin^2 2\theta_{e\mu}$ & $\sin^2 2\theta_{\mu\tau}$ &
$m_{\nu_\mu}/m_{\nu_e}$ & $m_{\nu_\tau}/m_{\nu_\mu}$ & $m_{\nu_\tau}$ [eV]\\
\hline
\multicolumn{6}{|c|}{$1_I$}\\
\hline
$1.5$ & 5.6\T{-2} & 2.8\T{-3} & 124 & 3100 & 6.2\\
$55$  & 5.2\T{-2} & 4.5\T{-3} & 124 & 1900 & 3.8\\
\hline
\multicolumn{6}{|c|}{$1_{II}$}\\
\hline
$1.5$ & 2.0\T{-2} & 2.8\T{-3} & 1100 & 1040 & 2.1\\
$55$  & 2.1\T{-2} & 4.5\T{-3} & 1100 & 650  & 1.3\\
\hline
\multicolumn{6}{|c|}{$2_I$}\\
\hline
$1.5$ & 1.9\T{-1} & 2.8\T{-2} & 33  & 6050 & 12\\
$55$  & 2.0\T{-1} & 5.4\T{-2} & 33  & 3750 & 7.5\\
\hline
\multicolumn{6}{|c|}{$2_{II}$}\\
\hline
$1.5$ & 2.8\T{-2} & 6.5\T{-4} & 2470 & 700 & 1.4\\
$55$  & 3.1\T{-2} & 1.7\T{-3} & 2460 & 435 & 0.87\\
\hline
\multicolumn{6}{|c|}{$4_I$}\\
\hline
$1.5$ & 3.6\T{-2} & 1.4\T{-3} & 434 & 1650 & 3.3\\
$55$  & 3.9\T{-2} & 2.2\T{-3} & 350 & 1150 & 2.3\\
\hline
\multicolumn{6}{|c|}{$4_{II}$}\\
\hline
$1.5$ & 2.4\T{-2} & 1.3\T{-2} & 2180 & 740 & 1.5\\
$55$  & 2.6\T{-2} & 2.0\T{-2} & 1500 & 555 & 1.1\\
\hline
\multicolumn{6}{|c|}{$4_{III}$}\\
\hline
$1.5$ & 1.7\T{-2} & 1.5\T{-3} & 2440 & 700 & 1.4\\
$55$  & 1.7\T{-2} & 2.3\T{-3} & 2100 & 470 & 0.94\\
\hline
\multicolumn{6}{|c|}{$4_{IV}$}\\
\hline
$1.5$ & 1.9\T{-2} & 1.3\T{-2} & 550  & 1480 & 3.0\\
$55$  & 1.8\T{-2} & 2.0\T{-2} & 655  & 840  & 1.68\\
\hline
\end{tabular}
\end{center}
\caption{The neutrino properties predicted by the ``good'' textures}
\label{RESULTS}
\end{table}

\begin{table}[h]
\begin{center}
\begin{tabular}{|l|l|l|}
\hline
up quarks & down quarks & charged leptons\\
\hline
$a_u = 1$ & $a_d = 1$ & \\
$b_u = \frac{2}{5}g_1^2 - (3 \lambda_b^2 + \lambda_\tau^2)$ &
$b_d = \frac{4}{5}g_1^2 - 3 \lambda_t^2$ & \\
$c_u = -2$ & $c_d = -2$ & \\
$d_u = -2$ & $d_d = -2$ & \\
$e_u = 0$ & $e_d = 0$ & \\
$x_1 = 3\lambda_t^2$ &
$x_6 = 3\lambda_b^2 +\lambda_\tau^2$ &
$x_{11} = 3\lambda_b^2 +\lambda_\tau^2$\\
$\quad- \left(\frac{13}{15}g_1^2 + 3g_2^2 +
\frac{16}{3}g_3^2\right)$ &
$\quad- \left(\frac{7}{15}g_1^2 + 3g_2^2 +\frac{16}{3}g_3^2\right)$ &
$\quad- \left(\frac{9}{5}g_1^2 + 3g_2^2 \right)$\\
$x_2 = 3$ & $x_7=3$ & $x_{12} = 3$\\
$x_3 = -9 y_t^4 - 3 y_t^2 y_b^2$ &
$x_8 = -9 y_b^4 - 3 y_t^2 y_b^2 - 3 y_\tau^4$&
$x_{13} = -9 y_b^4 - 3 y_t^2 y_b^2 - 3 y_\tau^4$\\
$\quad +y_t^2 (\frac{4}{5}g_1^2 + 16g_3^2)$ &
$\quad +y_b^2 (-\frac{2}{5} g_1^2 + 16g_3^2) + y_\tau^2(\frac{6}{5}g_1^2)$ &
$\quad +y_b^2 (-\frac{2}{5} g_1^2 + 16g_3^2) + y_\tau^2(\frac{6}{5}g_1^2)$\\
$\quad +\left(\frac{13}{15}(2n_g+\frac{3}{5})+\frac{169}{450}\right) g_1^4$ &
$\quad +\left(\frac{7}{15}(2n_g+\frac{3}{5})+\frac{49}{450}\right) g_1^4$ &
$\quad +\left(\frac{9}{5}(2n_g+\frac{3}{5})+\frac{81}{50}\right) g_1^4$\\
$\quad +\left(3(2n_g-5) + \frac{9}{2}\right)g_2^4$ &
$\quad +\left(3(2n_g-5) + \frac{9}{2}\right)g_2^4$ &
$\quad +\left(3(2n_g-5) + \frac{9}{2}\right)g_2^4$\\
$\quad +\left(\frac{16}{3}(2n_g-9)+\frac{128}{9}\right)g_3^4$ &
$\quad +\left(\frac{16}{3}(2n_g-9)+\frac{128}{9}\right)g_3^4$ &
$\quad +\frac{9}{5}g_1^2 g_2^2$\\
$\quad +g_1^2 g_2^2 +\frac{136}{45}g_1^2 g_3^2 + 8 g_2^2 g_3^2$ &
$\quad +g_1^2 g_2^2 +\frac{8}{9}g_1^2 g_3^2 + 8 g_2^2 g_3^2$ &
\\
$x_4 = \frac{2}{5}g_1^2 + 6g_2^2 - 9y_t^2$ &
$x_9 = \frac{4}{5}g_1^2 + 6g_2^2 - 9y_b^2 -3y_\tau^2$&
$x_{14} = 6g_2^2 - 9y_b^2 -3y_\tau^2$ \\
$x_5=-4$ &
$x_{10}=-4$ &
$x_{15}=-4$\\
\hline
\end{tabular}
\end{center}
\caption{RGE coefficients for the MSSM
\label{MSSMCoeff}}
\end{table}

\clearpage
\section*{FIGURE CAPTION}

Fig. I:
The dependence of $m_t$ on $\tan\beta$ for the texture $1_I$.

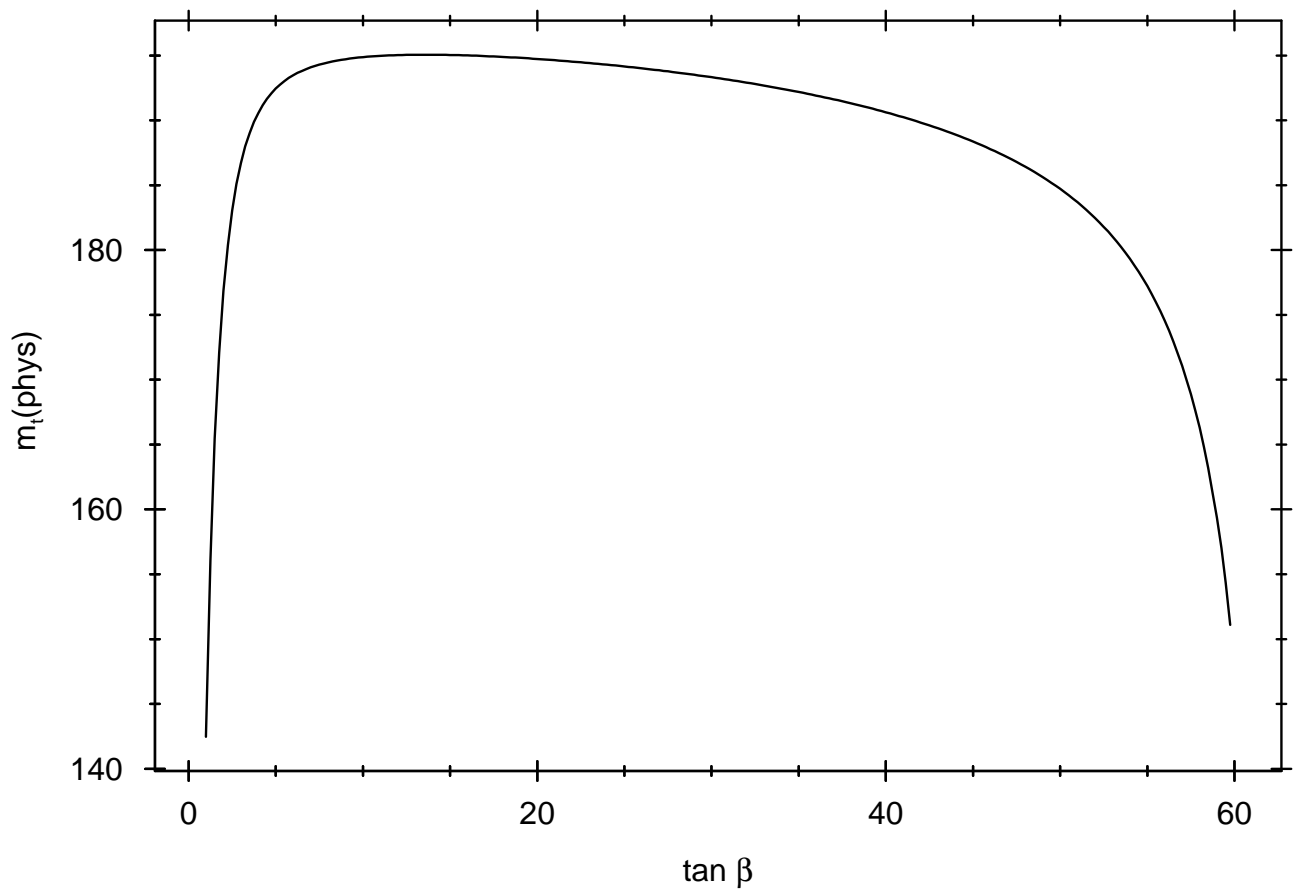
\end{document}